\documentclass[structabstract]{aa} 
\usepackage{graphicx}
\usepackage{txfonts}

\usepackage{natbib}
\bibpunct{(}{)}{;}{a}{}{,} 

\usepackage{acronym} 
\acrodef{2MASS}{Two Micron All Sky Survey}
\acrodef{BAT}{Burst Alert Telescope}
\acrodef{DAAD}{Deutscher Akademischer Austausch Dienst}
\acrodef{DFG}{Deutsche Forschungsgemeinschaft}
\acrodef{DSS}{Digitized Sky Survey}
\acrodef{ESO}{European Southern Observatory}
\acrodef{FRED}{fast rise and exponential decay}
\acrodef{GRB}{gamma-ray burst}
\acrodef{GROND}{Gamma-ray Burst Optical and Near-infrared Detector}
\acrodef{LMC}{Large Magellanic Cloud}
\acrodef{LSPO}{La Silla Paranal Observatory}
\acrodef{MPE}{Max-Planck-Institut f\"{u}r extraterrestrische Physik}
\acrodef{MPG}{Max-Planck-Gesellschaft}
\acrodef{MW}{Milky Way}
\acrodef{PSF}{point-spread function}
\acrodef{REFLEX}{ROSAT-ESO Flux Limited X-ray}
\acrodef{SB}{starburst}
\acrodef{SDSS}{Sloan Digital Sky Survey}
\acrodef{SED}{spectral energy distribution}
\acrodef{SMC}{Small Magellanic Cloud}
\acrodef{USNO}{United States Naval Observatory}
\acrodef{UVOT}{Ultraviolet and Optical Telescope}
\acrodef{VLT}{Very Large Telescope}
\acrodef{XRT}{X-ray Telescope}

\DeclareMathSymbol{\varGamma}{\mathord}{letters}{"00}
\DeclareMathSymbol{\varDelta}{\mathord}{letters}{"01}
\DeclareMathSymbol{\varTheta}{\mathord}{letters}{"02}
\DeclareMathSymbol{\varLambda}{\mathord}{letters}{"03}
\DeclareMathSymbol{\varXi}{\mathord}{letters}{"04}
\DeclareMathSymbol{\varPi}{\mathord}{letters}{"05}
\DeclareMathSymbol{\varSigma}{\mathord}{letters}{"06}
\DeclareMathSymbol{\varUpsilon}{\mathord}{letters}{"07}
\DeclareMathSymbol{\varPhi}{\mathord}{letters}{"08}
\DeclareMathSymbol{\varPsi}{\mathord}{letters}{"09}
\DeclareMathSymbol{\varOmega}{\mathord}{letters}{"0A}

\def \h {^\mathrm{h}} 
\def \m {^\mathrm{m}} 
\def \s {^\mathrm{s}} 

\def \fs {.\hspace{-0.25em}\s\hspace{-0.08em}} 

\def \d {^\circ}           
\def \am {^\prime}         
\def \as {^{\prime\prime}} 

\def \fas {.\hspace{-0.25em}\as\hspace{-0.08em}} 

\newcommand {\ECS} {\alpha, \delta\,\mathrm{(J2000.0)}}

\newcommand {\RA}[4] {#1\h #2\m #3\fs#4}
\newcommand {\Dec}[4] {#1\d #2\am #3\fas#4}

\newcommand {\oof}[2] {#1 \times 10^{#2}}

\def \Gg {g^\prime}
\def \Gr {r^\prime}
\def \Gi {i^\prime}
\def \Gz {z^\prime}
\def \GJ {J}
\def \GH {H}
\def \GK {K_s}

\newcommand {\TOe}[1] {$T_0$\,$=$\,#1}
\newcommand {\TOp}[1] {$T_0$\,$+$\,#1}
\newcommand {\TOm}[1] {$T_0$\,$-$\,#1}

\def \magAB {$\mathrm{mag}_\mathrm{AB}$}

\def \AV {A_V}

\def \bO {\beta_\mathrm{O}}
\def \aX {\alpha_\mathrm{X}}

\def \bOX {\beta_\mathrm{OX}}

\def \Xsqr {\chi^2}
\def \Xsqrred {\Xsqr_\mathrm{red.}}
\def \Xsqrdof {\Xsqr/\mathrm{d.o.f.}}

\def \Swift {\textit{Swift}}
\def \Fermi {\textit{Fermi}}

\def \Msun {\mathrm{M_\odot}}

\def \Ep {E_\mathrm{p}}

\def \Erest {E_{\gamma,\mathrm{iso.}}^\mathrm{rest}}

\begin{document}

\title{GRB\,071028B, a burst behind large amounts of dust in an unabsorbed galaxy}

\author{
C.~Clemens \inst{1} \thanks{\email{cclemens@mpe.mpg.de}}
\and
J.~Greiner \inst{1}
\and
T.~Kr\"{u}hler \inst{1,2}
\and
D.~Pierini \inst{1}
\and
S.~Savaglio \inst{1}
\and
S.~Klose \inst{3}
\and
P.~M.~J.~Afonso \inst{1}
\and
R.~Filgas \inst{1}
\and
F.~Olivares~E. \inst{1}
\and
A.~Rau \inst{1}
\and
P.~Schady \inst{1}
\and
A.~Rossi \inst{3}
\and
A.~K\"{u}pc\"{u}~Yolda\c{s} \inst{4}
\and
A.~C.~Updike \inst{5}
\and
A.~Yolda\c{s} \inst{1}
}
\authorrunning{C.~Clemens et al.}

\institute{
Max-Planck-Institut f\"{u}r extraterrestrische Physik, Giessenbachstra\ss{}e 1, 85748 Garching, Germany
\and
Universe Cluster, Technische Universit\"{a}t M\"{u}nchen, Boltzmannstra\ss{}e 2, 85748 Garching, Germany
\and
Th\"{u}ringer Landessternwarte Tautenburg, Sternwarte 5, 07778 Tautenburg, Germany
\and
European Southern Observatory, Karl-Schwarzschild-Stra\ss{}e 2, 85748 Garching, Germany
\and
Department of Physics and Astronomy, Clemson University, Clemson, SC 29634-0978, United States of America
}

\date{Received July~1, 2010; accepted January~12, 2011}

\abstract
{}
{We report on the discovery and properties of the fading afterglow and underlying host galaxy of GRB\,071028B, thereby facilitating a detailed comparison between these two.}
{Observations were performed with the Gamma-ray Burst Optical and Near-infrared Detector at the 2.2\,m telescope on the La Silla Paranal Observatory in Chile. We conducted five observations from 1.9\,d to 227.2\,d after the trigger and obtained deep images in the $\Gg\Gr\Gi\Gz$ and $\GJ\GH\GK$ bands.}
{Based on accurate seven-channel photometry covering the optical to near-infrared wavelength range, we derive a photometric redshift of $z = 0.94\,^{+0.05}_{-0.10}$ for the unabsorbed host galaxy of GRB\,071028B. In contrast, we show that the afterglow with an intrinsic extinction of $\AV^\mathrm{SB} = (0.70 \pm 0.11)\,\mathrm{mag}$ is moderately absorbed and requires a relatively flat extinction curve. According to the reported \Swift/BAT observations, the energetics yield an isotropic energy release of $\Erest = \oof{(1.4\,^{+2.4}_{-0.7})}{51}\,\mathrm{erg}$.}
{}

\keywords{Gamma-rays: bursts: individual: GRB\,071028B -- Galaxies: starburst -- Galaxies: photometry -- Techniques: photometric}

\maketitle

\section{Introduction}


\Swift\ is primarily designed for the rapid and precise localisation of \acp{GRB} and their afterglow \citep{swift}. Utilising the on-board \ac{BAT} \citep{bat}, the satellite continuously surveys an area in the sky of about two steradians and reacts to increased gamma-ray emission. Once a \ac{GRB} has been detected, \Swift\ determines an initial position to a precision of about $3\am$ and autonomously slews to the new location within a few seconds. The \ac{XRT} \citep{xrt}, which is a second on-board instrument with a narrow field-of-view, now points towards the \ac{GRB}, yielding a refined position measurement with an accuracy of about $3\as$ -- $5\as$. Simultaneously, the on-board \ac{UVOT} \citep{uvot} is utilised for further improvement of the \ac{GRB} location.
\Swift\ thus provides an accurate position within a minute after the trigger and ground-based follow-up observations can be promptly initiated.

However, there are several factors that prevent the automated detection of \acp{GRB} by \Swift: weak, short-, or very long-duration bursts, bursts located near the edge of the field-of-view, and bursts occurring shortly before or during a preplanned slew manoeuvre. Therefore, the data are routinely streamed to the ground and attempts are made by the \Swift/BAT team to detect any missed \acp{GRB}.
Before the late ground-based detection of GRB\,071028B, only eight (3\%) similar cases out of 277 \acp{GRB} detected by \Swift\ had been known: GRB\,051012 \citep{gcn4093}, GRB\,051114 \citep{gcn4272}, GRB\,060505 \citep{gcn5076}, GRB\,070227 \citep{gcn6156}, GRB\,070326 \citep{gcn6653}, GRB\,070406 \citep{gcn6247}, GRB\,071006 \citep{gcn6858}, and GRB\,071010C \citep{gcn6906}. In these cases, the average time delay between the \ac{GRB} and the reported localisation is 34\,h.

Before we report on our observations performed with the \ac{GROND} and our analysis of the afterglow and host galaxy of GRB\,071028B, we review the corresponding \Swift\ observations.

\section{\Swift\ observations of GRB\,071028B}


GRB\,071028B (trigger \#295492) was observed by the \Swift/BAT on October~28, 2007, at \TOe{02:43:46\,UT}. Since this \ac{GRB} was located near the edge of the field-of-view, no source was immediately found on-board, so no prompt \Swift/XRT and \Swift/UVOT observations were performed. The late ground-based analysis of the \Swift/BAT data revealed a source at the coordinates of $\ECS = 23\h 36\m 39\s, -31\d 37\am 48\as$ with an error radius of $3\am$, where this position was announced with a time delay of 58.6\,h \citep{gcn7019}.

\Swift/XRT observations of GRB\,071028B were scheduled and performed on October~30, 2007, at 13:36\,UT (\TOp{2.5\,d}) for 14.1\,ks, detecting one known X-ray source with a likely optical counterpart from the \ac{USNO} catalogue, and two rather faint, uncatalogued X-ray sources within the \Swift/BAT error circle. No optical counterparts of the unknown X-ray sources were discovered with the \Swift/UVOT to a $3\sigma$ upper limit of $m_V = 21.3\,\mathrm{mag}$ in a total of 6443\,s exposure \citep{gcn7040}.

On November~7, 2007, at 00:02\,UT (\TOp{9.9\,d}) and on November~10, 2007, at 06:44\,UT (\TOp{13.2\,d}) follow-up observations were performed with the \Swift/XRT for 9\,ks and 11\,ks, respectively. One of the two faint objects located at the coordinates of $\ECS = \RA{23}{36}{39}{12}, \Dec{-31}{37}{13}{8}$ with an error radius of $4\fas8$ had clearly faded below a $3\sigma$ upper limit of $\oof{6.7}{-4}\,\mathrm{cts}\,\mathrm{s}^{-1}$, indicating that this was the X-ray afterglow of GRB\,071028B \citep{gcn7064}.
\cite{but07} determines the astrometrically corrected \Swift/XRT position of \acp{GRB}. As a result, the X-ray afterglow coordinates of GRB\,071028B were revised to $\ECS = \RA{23}{36}{38}{67}, \Dec{-31}{37}{17}{0}$ with an error radius of $3\fas4$.

\Swift/BAT observations of GRB\,071028B in the 15\,keV -- 150\,keV energy range yielded a light curve showing two \ac{FRED} peaks at \TOp{1\,s} and \TOp{48\,s} with a duration of about 4\,s and 9\,s, respectively \citep{gcn7019}. Integrating this light curve from \TOm{0.5\,s} and over a time interval of 1\,s, \cite{gcnr105} determined a peak photon flux of $\varPhi = (1.4 \pm 0.5)\,\mathrm{ph}\,\mathrm{cm}^{-2}\,\mathrm{s}^{-1}$. This \ac{GRB} released 90\% of its total energy within a time period of $T_{90} = (55 \pm 2)\,\mathrm{s}$ \citep{gcnr105}, thereby falling into the class of long-duration bursts \citep{kou93}.

The \Swift/XRT spectrum of GRB\,071028B covering the 0.3\,keV -- 10\,keV energy range can be described by a photon index of $\varGamma = 2.0\,^{+1.1}_{-0.8}$ \citep{eva09} and an intrinsic hydrogen column density consistent with the Galactic value of $N_\mathrm{H} = \oof{1.2}{20}\,\mathrm{cm}^{-2}$ \citep{kal05}.

All quoted errors deduced from \Swift\ observations are at the 90\% confidence level.

\section{GROND observations of GRB\,071028B}

\ac{GROND} is mounted at the 2.2\,m telescope of the \ac{MPG}, which is operated by the \ac{ESO} on the \ac{LSPO} in Chile, and is capable of simultaneous optical and near-infrared imaging \citep{grond}. The instrument typically takes series of 46\,s, 137\,s, or 408\,s integrations in the $\Gg\Gr\Gi\Gz$ bands with gaps of about 70\,s. In parallel, the three near-infrared detectors ($\GJ\GH\GK$ bands) perform 10\,s integrations, which are separated by about 6\,s. These unavoidable gaps are caused by read-out, data transfer, telescope presets, and dithering.

\begin{table}[ht]
\caption{Observation log of GRB\,071028B.}
\label{tab_obs}
\centering
\begin{tabular}{llll}
\hline\hline
\noalign{\smallskip}
\# & Date & Time [UT] & $T-T_0$ [d] \\
\noalign{\smallskip}
\hline
\noalign{\smallskip}
1 & Oct.~30, 2007 & 00:24 -- 03:17 & $+$\,1.9 \\
\noalign{\smallskip}
2 & Nov.~12, 2007 & 00:22 -- 02:31 & $+$\,14.9 \\
\noalign{\smallskip}
3 & Dec.~1, 2007 & 00:40 -- 03:03 & $+$\,33.9 \\
\noalign{\smallskip}
4 & Jun.~5, 2008 & 09:49 -- 10:32 & $+$\,221.3 \\
\noalign{\smallskip}
5 & Jun.~11, 2008 & 07:45 -- 10:16 & $+$\,227.2 \\
\noalign{\smallskip}
\hline
\end{tabular}
\tablefoot{The start time $T$ of each GROND observation is given relative to the time $T_0$ of the \Swift/BAT trigger.}
\end{table}

\begin{figure}[th]
\includegraphics[width=\hsize]{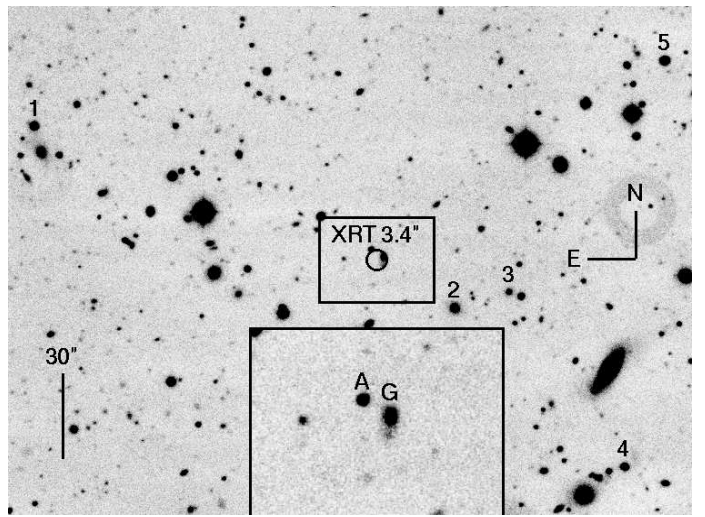} 
\caption[fc]{Finding chart covering the $4\am \times 3\am$ field of GRB\,071028B as observed with GROND for 2.22\,h in the $\Gr$ band during the first observing night. The circle represents the astrometrically corrected \Swift/XRT position with its 90\% error radius. All secondary standard stars used for the photometric calibration (see Table~\ref{tab_stds}), the afterglow (A), and a putative foreground galaxy (G) are labelled. A scale-up of the afterglow area is shown in the bottom panel.}
\label{fig_fc}
\end{figure}

\begin{figure}[th]
\includegraphics[width=\hsize]{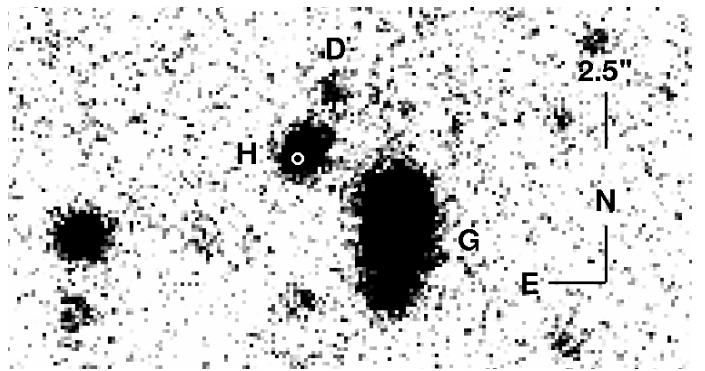} 
\caption[fc]{Scale-up showing the suspected host galaxy (H) of GRB\,071028B, a putative foreground galaxy (G), and a potential dwarf galaxy (D) as observed with GROND for 5.7\,h in the $\Gr$ band. The small circle represents the position of the afterglow. Several images taken at \TOp{14.9\,d} and later are combined to maximise the signal-to-noise ratio.}
\label{fig_host}
\end{figure}

\ac{GROND} observations of GRB\,071028B could already be initiated at \TOp{45.7\,h}, since we received a private communication announcing a preliminary \Swift/BAT position \citep{gre07}. The log of all five observations is listed in Table~\ref{tab_obs}. In addition, the finding chart is displayed in Fig.~\ref{fig_fc} and a scale-up of GRB\,071028B and the surrounding area is shown in Fig.~\ref{fig_host}.

\section{Data reduction and analysis}

The data reduction and analysis was done in a standard way by applying a bad pixel map, dark and bias frames, a flat-field and fringe correction ($\Gz$ band) to each raw image. We utilised the \textit{GROND Manual Analysis Pipeline v1.6} \citep{gp}, which is mainly based on commands from the \textit{IRAF} package \citep{iraf} and \textit{SExtractor} \citep{sextractor}.

All exposures of the first observing night within a single filter band were stacked to maximise the signal-to-noise ratio of the afterglow, yielding effective exposure times of 2.22\,h and 1.73\,h for the $\Gg\Gr\Gi\Gz$ and $\GJ\GH\GK$ bands, respectively.
We combined the corresponding frames of the second, third, and fifth observing night, obtaining deep images of the surrounding galaxies with effective exposure times of 5.71\,h and 4.32\,h for the $\Gg\Gr\Gi\Gz$ and $\GJ\GH\GK$ bands, respectively. The images of the fourth observing night were excluded from the stacking, since they were taken under bad seeing conditions.

Separately for the photometry of the afterglow and the surrounding galaxies, the corresponding seven images were first aligned to the $\Gg$ band frame using five isolated stars to determine their relative shift, rotation and scale. The \acp{PSF} of these images were then matched to the worst \ac{PSF} ($\GK$ band), since the average seeing shows intolerable variations from $1\fas12$ to $1\fas65$ and $1\fas20$ to $1\fas69$ for the individual filter bands of both data sets, respectively.

Running \textit{SExtractor} in \texttt{Dual\,Mode} and using the deepest image ($\Gr$ band) as the reference frame, the photometry was extracted by applying the same apertures, which were fixed in size and position, to the remaining images. The detection routine was set to recognise sources consisting of at least two pixels, which are $3\sigma$ above the background, and a minimum contrast parameter of 0.002 with 32 subthresholds for deblending.

\begin{table*}[ht]
\caption{Apparent brightnesses (in the AB system) and errors for the secondary photometric standard stars of GRB\,071028B.}
\label{tab_stds}
\centering
\begin{tabular}{lllllllll}
\hline\hline
\noalign{\smallskip}
\# & $\ECS$ & $m_{\Gg}$ [\magAB] & $m_{\Gr}$ [\magAB] & $m_{\Gi}$ [\magAB] & $m_{\Gz}$ [\magAB] & $m_{\GJ}$ [\magAB] & $m_{\GH}$ [\magAB] & $m_{\GK}$ [\magAB] \\
\noalign{\smallskip}
\hline
\noalign{\smallskip}
1 &
$\RA{23}{36}{48}{01}, \Dec{-31}{36}{29}{9}$ &
$20.11 \pm 0.09$ & $20.08 \pm 0.02$ & $20.18 \pm 0.04$ & $20.03 \pm 0.05$ &
$19.36 \pm 0.08$ & $19.40 \pm 0.09$ & $18.88 \pm 0.09$ \\
\noalign{\smallskip}
2 &
$\RA{23}{36}{36}{54}, \Dec{-31}{37}{33}{9}$ &
$22.47 \pm 0.10$ & $21.00 \pm 0.03$ & $20.29 \pm 0.05$ & $19.91 \pm 0.06$ &
$19.39 \pm 0.08$ & $18.99 \pm 0.09$ & $18.47 \pm 0.09$ \\
\noalign{\smallskip}
3 &
$\RA{23}{36}{35}{05}, \Dec{-31}{37}{28}{3}$ &
$23.86 \pm 0.13$ & $22.04 \pm 0.04$ & $20.81 \pm 0.05$ & $20.33 \pm 0.06$ &
$19.70 \pm 0.09$ & $19.17 \pm 0.09$ & $18.74 \pm 0.09$ \\
\noalign{\smallskip}
4 &
$\RA{23}{36}{31}{90}, \Dec{-31}{38}{29}{8}$ &
$22.01 \pm 0.09$ & $20.53 \pm 0.03$ & $19.93 \pm 0.04$ & $19.61 \pm 0.05$ &
$19.15 \pm 0.08$ & $18.74 \pm 0.09$ & $18.32 \pm 0.09$ \\
\noalign{\smallskip}
5 &
$\RA{23}{36}{30}{80}, \Dec{-31}{36}{06}{9}$ &
$21.61 \pm 0.09$ & $20.39 \pm 0.03$ & $19.86 \pm 0.05$ & $19.52 \pm 0.05$ &
$19.12 \pm 0.08$ & $18.72 \pm 0.09$ & $18.43 \pm 0.09$ \\
\noalign{\smallskip}
\hline
\end{tabular}
\end{table*}

Standard \ac{PSF} photometry was applied to conduct the photometric calibration.
\ac{GROND} observed the field of GRB\,071028B on June~15, 2008, at 08:45\,UT (\TOp{231.3\,d}) for 2.4\,min in the $\Gg\Gr\Gi\Gz$ bands and 4.0\,min in the $\GJ\GH\GK$ bands under photometric conditions and at low airmass. Shortly afterwards at 08:54\,UT (\TOp{231.3\,d}), the nearby field of standard star SA\,114-750 \citep{smi02} was imaged for 12\,s and 48\,s, respectively. The latter observation allowed us to derive zero points for the $\Gg\Gr\Gi\Gz$ and $\GJ\GH\GK$ bands by comparing catalogue stars from the \ac{SDSS} \citep{sdss} and the \ac{2MASS} \citep{2mass} with SA\,114-750 field stars. Subsequently, bright stars with well-sampled \acp{PSF} in all seven filter bands were chosen as close as possible to GRB\,071028B. Based on the calibrated zero points, their magnitudes were then transformed to the AB system \citep{absystem}. In this way, a set of five secondary photometric standard stars for relative photometry was established (see Table~\ref{tab_stds}).
The photometry of the afterglow and the surrounding galaxies was corrected assuming a Galactic extinction of $\AV = 0.06\,\mathrm{mag}$ \citep{sch98}.
The astrometry was obtained by matching the positions of field stars in $\Gr$ band frames with their coordinates listed in the \ac{USNO} catalogue \citep{mon03}. Including the statistical error of the astrometric solution, the overall precision is approximately $0\fas3$.

We have adopted a flat $\Lambda$-dominated universe with a normalised cosmological constant of $\varOmega_\Lambda = 0.73$, a matter density of $\varOmega_\mathrm{m} = 0.27$ and a Hubble constant of $H_0 = 71\,\mathrm{km}\,\mathrm{s}^{-1}\,\mathrm{Mpc}^{-1}$ throughout the paper \citep{spe03}.
Cosmological parameters (e.g. angular scale) were computed using the \textit{Cosmology Calculator} coded by \cite{coscalc}.
The properties of the afterglow and the surrounding galaxies (e.g. absolute brightness) were estimated using \textit{HyperZ v11} originally developed by \cite{hyperz}.
The analysis of the broad-band \ac{SED} was done using the numerical minimisation program \textit{Minuit} \citep{minuit}, which is included in the software package \textit{ROOT v5.27.06} \citep{root} and was modified to fit the needs of \ac{GRB} afterglow physics.
All errors estimated with \textit{HyperZ} and \textit{Minuit} are given at the $1\sigma$ confidence level.

\section{Results and discussion of GRB\,071028B}

In the images of the first observing night, a single point source (A) was detected at the coordinates of $\ECS = \RA{23}{36}{38}{83}, \Dec{-31}{37}{13}{3}$ with an uncertainty of $0\fas3$. This is within the original \Swift/XRT error circle \citep{gcn7064}, but just outside of the error circle determined by \cite{but07}. The point source had also clearly faded in later images, indicating that this was the optical and near-infrared afterglow of GRB\,071028B (see Sect.~\ref{sec_res_aglow}).

Starting with the second observing night at \TOp{14.9\,d}, an extended object (H) next to the afterglow position was clearly detected in the images. Based on an angular scale of $7.9\,\mathrm{kpc}/\as$ derived from a photometric redshift of $z = 0.94\,^{+0.05}_{-0.10}$ for this constant source, the angular offset of $0\fas3\,^{+0\fas4}_{-0\fas3}$ between the afterglow and this object corresponds to a projected distance of $(2.7\,^{+2.9}_{-2.7})\,\mathrm{kpc}$. This is twice as large as the median projected distance of 1.3\,kpc found by \cite{blo02b} for a sample of 20 hosts of long-duration \acp{GRB}, but fully consistent within the errors. We therefore interpret this object as the host galaxy of GRB\,071028B, which is centred on the coordinates of $\ECS = \RA{23}{36}{38}{81}, \Dec{-31}{37}{13}{1}$ with an error of $0\fas3$ (see Sect.~\ref{sec_res_host}).

Southwest of the afterglow position, there is a prominent and extended object (G) located at the central coordinates of $\ECS = \RA{23}{36}{38}{49}, \Dec{-31}{37}{16}{1}$ with an uncertainty of $0\fas3$. The angular offset of $6\fas4 \pm 0\fas4$ between the afterglow and this constant source at a photometric redshift of $z = 0.54\,^{+0.09}_{-0.10}$ corresponds to a projected distance of $(41 \pm 2)\,\mathrm{kpc}$. Since this is 31 times larger than the median projected distance of \ac{GRB} hosts, we conclude that this object is not associated with GRB\,071028B and instead represents a foreground galaxy (see Sect.~\ref{sec_res_galaxy}).

The extended object (D) northwest of the afterglow position is centred on the coordinates of $\ECS = \RA{23}{36}{38}{70}, \Dec{-31}{37}{10}{5}$ with an error of $0\fas3$. We now assume that this object represents a host galaxy at an unknown redshift $z$. In this way, we estimate a lower limit of $z \approx 0.7$ by adopting the lowest redshift in the sample of hosts of long-duration \acp{GRB}, which are fainter than this constant source with an apparent brightness of $m_{\Gr} \approx 24.4\,\mathrm{mag}$ \citep{sav09}. We then calculate a minimum projected distance of 24\,kpc between the afterglow and this object, thus making an association of this source with GRB\,071028B very implausible. Given the detection in only $\Gg$ and $\Gr$ band images, we suppose this is a very blue and young dwarf galaxy.

\begin{table*}[ht]
\caption{Apparent brightnesses (in the AB system) and errors for the afterglow (A) and host galaxy (H) of GRB\,071028B.}
\label{tab_mags}
\centering
\begin{tabular}{lllllllll}
\hline\hline
\noalign{\smallskip}
\# & $\ECS$ & $m_{\Gg}$ [\magAB] & $m_{\Gr}$ [\magAB] & $m_{\Gi}$ [\magAB] & $m_{\Gz}$ [\magAB] & $m_{\GJ}$ [\magAB] & $m_{\GH}$ [\magAB] & $m_{\GK}$ [\magAB] \\
\noalign{\smallskip}
\hline
\noalign{\smallskip}
A &
$\RA{23}{36}{38}{83}, \Dec{-31}{37}{13}{3}$ &
$22.78 \pm 0.13$ & $22.33 \pm 0.07$ & $21.91 \pm 0.08$ & $21.49 \pm 0.08$ &
$20.98 \pm 0.11$ & $20.44 \pm 0.11$ & $19.63 \pm 0.12$ \\
\noalign{\smallskip}
H &
$\RA{23}{36}{38}{81}, \Dec{-31}{37}{13}{1}$ &
$23.59 \pm 0.12$ & $23.60 \pm 0.09$ & $23.00 \pm 0.08$ & $22.75 \pm 0.10$ &
$22.24 \pm 0.12$ & $21.95 \pm 0.13$ & $21.70 \pm 0.14$ \\
\noalign{\smallskip}
G &
$\RA{23}{36}{38}{49}, \Dec{-31}{37}{16}{1}$ &
$22.63 \pm 0.12$ & $21.54 \pm 0.09$ & $20.94 \pm 0.08$ & $20.73 \pm 0.10$ &
$20.29 \pm 0.12$ & $19.97 \pm 0.10$ & $19.60 \pm 0.11$ \\
\noalign{\smallskip}
\hline
\end{tabular}
\tablefoot{The values of the foreground galaxy (G) are listed in the last row. The afterglow was observed on October~30, 2007, at 01:50\,UT (\TOp{2.0\,d}).}
\end{table*}

\begin{figure*}[th]
\sidecaption
\includegraphics[width=120mm]{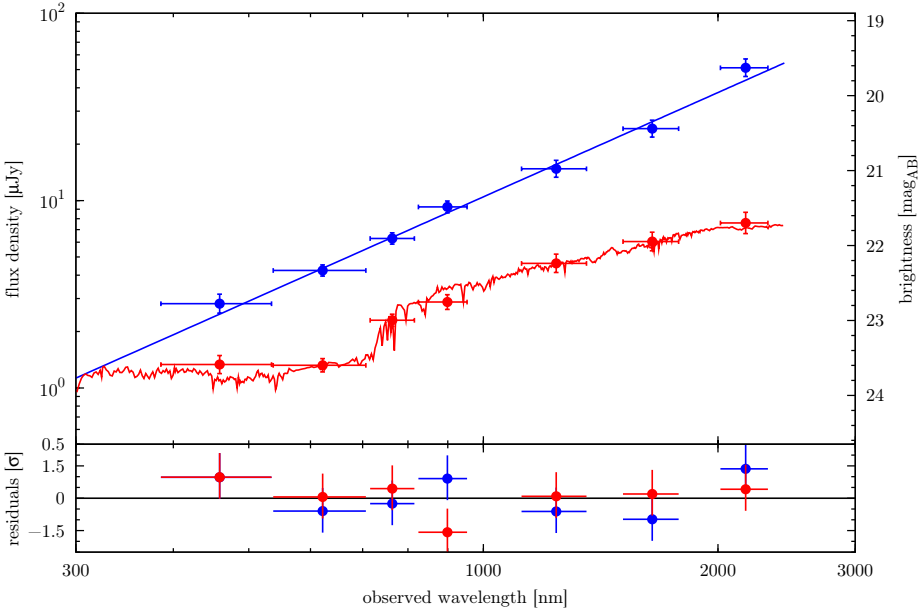} 
\caption[fc]{Optical and near-infrared spectral energy distributions of the afterglow (top) taken on October~30, 2007, at 01:50\,UT (\TOp{2.0\,d}) and host galaxy (bottom) of GRB\,071028B. Using \textit{HyperZ}, the data points are best fit by a single power law with a spectral index of $\bO^\mathrm{none} = 1.85\,^{+0.09}_{-0.08}$ (see Table~\ref{tab_fit_aglow}) and a model of an Sb galaxy with a photometric redshift of $z = 0.94\,^{+0.04}_{-0.05}$ (see Table~\ref{tab_fit_host}).}
\label{fig_mags}
\end{figure*}

\subsection{Properties of the host galaxy}
\label{sec_res_host}

The photometry of the host galaxy (H) yields the apparent brightnesses presented in Table~\ref{tab_mags} and Fig.~\ref{fig_mags}.


We utilised \textit{HyperZ} for fitting the model of a starburst (SB), as well as models of elliptical (E), irregular (I), and spiral (Sx) galaxies to the optical and near-infrared \ac{SED} of the host galaxy \citep{bru03}.
Besides the simplest case of an unabsorbed host galaxy, we examined scenarios based on dust models of the \ac{MW} \citep{sea79}, the \ac{LMC} \citep{fit86}, the \ac{SMC} \citep{pre84,bou85}, and a \ac{SB} \citep{cal00}. These \ac{MW}, \ac{LMC}, and \ac{SMC} extinction laws are strictly valid for individual lines-of-sight, whereas the \ac{SB} attenuation law, which is likely to be associated with the \ac{SMC} extinction law, considers scattering and propagation in extended objects.

During the fitting, we limited the photometric redshift $z$ of the host galaxy to a parameter range of $0.02 \leq z \leq 1.20$ (steps of 0.01), since no \ac{GRB} host with apparent brightnesses of $m_R \leq 23.9\,\mathrm{mag}$ and $m_K \leq 23.3\,\mathrm{mag}$ is known at a redshift of $z \geq 1.2$ \citep{sav09}. Moreover, the clearly detected emission of the afterglow in the $\Gg$ band signifies that the Lyman break is lying bluewards of the shortest filter band, implying a redshift of $z \leq 3.3$. The age $\tau$ of the dominant stellar population was allowed to vary between a minimum value of 1.0\,Myr and the age of the Universe at the fitted redshift. A wide parameter range of $0.00\,\mathrm{mag} \leq \AV \leq 1.60\,\mathrm{mag}$ (steps of $0.04\,\mathrm{mag}$) was allowed for the intrinsic extinction $\AV$.

\begin{table}[ht]
\caption{Results for the host galaxy of GRB\,071028B.}
\label{tab_fit_host}
\centering
\begin{tabular}{llll}
\hline\hline
\noalign{\smallskip}
Model & $z$ & $\tau$ [Gyr] & $\Xsqrred$ \\
\noalign{\smallskip}
\hline
\noalign{\smallskip}
SB & 0.06\,$^{+0.99}_{-0.04}$ & 11.5\,$^{+2.0}_{-11.5}$ & 10.1 \\
\noalign{\smallskip}
E & 0.96\,$^{+0.04}_{-0.03}$ & 2.0\,$^{+0.3}_{-0.3}$ & 1.86 \\
\noalign{\smallskip}
I & 0.94\,$^{+0.04}_{-0.10}$ & 6.5\,$^{+2.0}_{-1.0}$ & 1.76 \\
\noalign{\smallskip}
S0 & 0.96\,$^{+0.03}_{-0.06}$ & 3.5\,$^{+1.0}_{-1.2}$ & 1.51 \\
\noalign{\smallskip}
Sa & 0.96\,$^{+0.03}_{-0.05}$ & 3.5\,$^{+2.0}_{-0.9}$ & 1.14 \\
\noalign{\smallskip}
Sb & 0.94\,$^{+0.04}_{-0.05}$ & 5.5\,$^{+1.0}_{-2.0}$ & 0.89 \\
\noalign{\smallskip}
Sc & 0.94\,$^{+0.04}_{-0.08}$ & 6.5\,$^{+2.0}_{-1.0}$ & 0.99 \\
\noalign{\smallskip}
Sd & 0.94\,$^{+0.04}_{-0.10}$ & 6.5\,$^{+2.0}_{-1.0}$ & 1.32 \\
\noalign{\smallskip}
\hline
\end{tabular}
\end{table}

Table~\ref{tab_fit_host} shows the simplest scenarios without dust in the host galaxy,
where the scenarios based on the model of an elliptical galaxy, an irregular galaxy, or a starburst are implausible.
In contrast, the best-fit solution with a goodness-of-fit of $\Xsqrdof = 3.54/4$ is based on the model of an Sb galaxy,
but it is impossible to distinguish this model from those of other spiral galaxies.
We also note that these good-fit solutions yield comparable redshifts of $z \approx 0.94$.
Applying dust models and thereby introducing the intrinsic extinction $\AV$ as an additional free parameter does not significantly improve the fits, since fitting these scenarios shows only marginally better $\Xsqr$-statistics ($\Xsqrdof \geq 2.91/3$). Moreover, similar results to the scenarios without dust are obtained in this way.
We therefore adopt a photometric redshift of $z = 0.94\,^{+0.05}_{-0.10}$ and an age $\tau$ of the dominant stellar population between 2.3\,Gyr and 8.5\,Gyr for the host galaxy.

Assuming that the prominent colour-term of $m_{\Gr} - m_{\Gi} \approx 0.6\,\mathrm{mag}$ between the $\Gr$ and $\Gi$ bands indicates the position of both the Balmer break and the Ca\,II H\&K doublet, we independently estimate a redshift of $z \approx 0.9$, which is consistent with the photometric value.
The estimated apparent brightnesses of $m_{\Gr} = (23.60 \pm 0.09)\,\mathrm{mag}$ and $m_{\GK} = (21.70 \pm 0.14)\,\mathrm{mag}$ for the host galaxy, as well as its absolute brightness of $M_K \approx -21.5\,\mathrm{mag}$ are not exceptional, and agree with the sample of \ac{GRB} hosts compiled by \cite{sav09}.

\cite{sav09} have mainly studied \ac{GRB} hosts at a redshift of $z \leq 1.6$, and establish the following correlation between the stellar mass $M_\ast$ and the dust-corrected absolute brightness $M_K$ (in the AB system):
$$\log M_\ast = -0.467 \, M_K - 0.179,$$
where this formula is valid provided that the Balmer break is located bluewards of the $K$ band. We note that the stellar mass $M_\ast$ is subject to a minimum dispersion of roughly a factor of two. By applying the absolute brightness of $M_K \approx -21.5\,\mathrm{mag}$ to the equation above, we calculate a stellar mass of $M_\ast \approx \oof{7.0}{9}\,\Msun$ for the host galaxy. This agrees with the stellar mass of $M_\ast \approx \oof{31}{9}\,\Msun$ derived from the model fitting, since both procedures are affected by a large uncertainty. We therefore conclude that the host galaxy belongs to the category of dwarf galaxies. In comparison, we derive a lower median stellar mass of about $\oof{2.0}{9}\,\Msun$ for hosts of long-duration \acp{GRB} from the results provided by \cite{sav09}.

\subsection{Properties of the foreground galaxy}
\label{sec_res_galaxy}

We assume for the photometry of the foreground galaxy (G) that the clearly visible bulge and the extended structure south of it constitute a single irregular galaxy. In this way, we estimate the apparent brightnesses presented in Table~\ref{tab_mags}.



We applied the same methods as already used for analysing the host galaxy \ac{SED} on the optical and near-infrared \ac{SED} of the foreground galaxy, and calculate a minimum goodness-of-fit of $\Xsqrdof \geq 4.00/4$ for the scenarios without dust. In contrast, the best-fit solution with a superior goodness-of-fit of $\Xsqrdof = 0.668/3$ is obtained by applying dust models. According to the theorem of \cite{wil38}, this improvement of $\Delta\Xsqr = 3.33$ justifies introducing the intrinsic extinction $\AV$ as an additional free parameter with a probability of 93\%.
In this way, the best-fit solution yields a photometric redshift of $z = 0.54\,^{+0.09}_{-0.10}$ for the foreground galaxy consistent with the photometric redshift $z$ of the scenarios without dust.

This lower redshift compared to the host galaxy is also demonstrated by both the prominent Balmer break and the Ca\,II H\&K doublet between the $\Gg$ and $\Gr$ bands ($m_{\Gg} - m_{\Gr} \approx 1.1\,\mathrm{mag}$).

\subsection{Properties of the afterglow}
\label{sec_res_aglow}


The photometry of the afterglow (A) is obtained by eliminating the coexistent emission of the host galaxy, since the latter contributes up to 32\% ($\Gg$ band) to the entire light.
We therefore subtract the total fluxes of the host galaxy from the corresponding total fluxes of the afterglow, and propagate the errors accordingly.
This approach is preferred to the extraction of the afterglow photometry from a residual image since the uncertainty of the recovered afterglow flux is minimised.
In this way, we secure an unbiased photometry of the afterglow and estimate the apparent brightnesses presented in Table~\ref{tab_mags} and Fig.~\ref{fig_mags}.


Utilising a modified version of \textit{HyperZ}, we fitted a single power law to the optical and near-infrared \ac{SED} of the afterglow.
Besides the simplest case of an unabsorbed afterglow, we examined scenarios based on the same dust models as already used for analysing the host galaxy \ac{SED}.

During the fitting, we fixed the redshift $z$ of the afterglow to the photometric redshift $z$ of the host galaxy ($z = 0.94$), since the latter is already known from an independent fit. We further limited the spectral index $\bO$ and the intrinsic extinction $\AV$ to a parameter range of $0.00 \leq \bO \leq 2.20$ (steps of 0.01) and $0.00\,\mathrm{mag} \leq \AV \leq 1.60\,\mathrm{mag}$ (steps of $0.04\,\mathrm{mag}$), respectively.

\begin{table}[ht]
\caption{Results for the afterglow of GRB\,071028B.}
\label{tab_fit_aglow}
\centering
\begin{tabular}{lllll}
\hline\hline
\noalign{\smallskip}
Dust Model & $z$ & $\bO$ & $\AV$ [mag] & $\Xsqrred$ \\
\noalign{\smallskip}
\hline
\noalign{\smallskip}
none & -- & 1.85\,$^{+0.09}_{-0.08}$ & -- & 0.98 \\
\noalign{\smallskip}
MW & 0.94 & 1.85\,$^{+0.09}_{-0.13}$ & 0.00\,$^{+0.12}_{-0.00}$ & 1.22 \\
\noalign{\smallskip}
LMC & 0.94 & 1.85\,$^{+0.09}_{-0.14}$ & 0.00\,$^{+0.12}_{-0.00}$ & 1.22 \\
\noalign{\smallskip}
SMC & 0.94 & 1.85\,$^{+0.09}_{-0.18}$ & 0.00\,$^{+0.16}_{-0.00}$ & 1.22 \\
\noalign{\smallskip}
SB & 0.94 & 1.85\,$^{+0.09}_{-0.38}$ & 0.00\,$^{+0.40}_{-0.00}$ & 1.22 \\
\noalign{\smallskip}
\hline
\end{tabular}
\end{table}

Table~\ref{tab_fit_aglow} shows the details of the fits. For the simplest scenario without dust in the afterglow, the best-fit solution reaches a goodness-of-fit of $\Xsqrdof = 4.87/5$. Moreover, the remaining scenarios based on dust models are best fit with equal $\Xsqr$-statistics and identical results compared to the scenario without dust. We therefore note that the respective dust models cannot be distinguished.
In this way, the best-fit solution yields a spectral index of $\bO^\mathrm{none} = 1.85\,^{+0.09}_{-0.08}$ for the afterglow.

\begin{figure*}[th]
\sidecaption
\includegraphics[width=120mm]{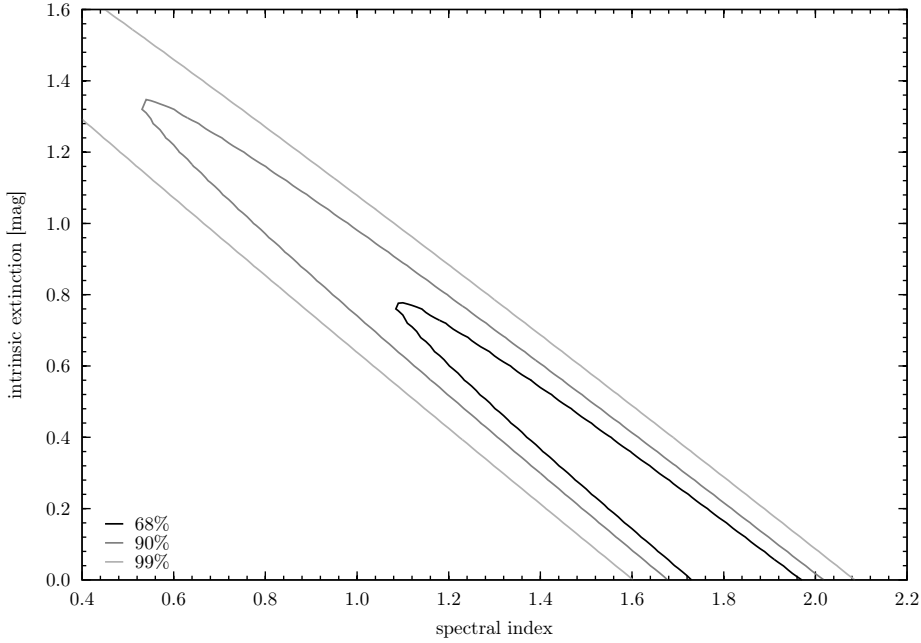} 
\caption[fc]{Contour plot of the afterglow of GRB\,071028B obtained by fitting models to its optical and near-infrared spectral energy distribution using \textit{HyperZ}. The figure shows the 68\%, 90\%, and 99\% confidence levels of the spectral index $\bO^\mathrm{SB}$ versus the intrinsic extinction $\AV^\mathrm{SB}$, which are based on a single power law in combination with the dust model of a starburst (SB).}
\label{fig_fit_aglow}
\end{figure*}

However, the spectral index $\bO^\mathrm{none}$ is much higher than expected from the \Swift/XRT observations of \acp{GRB} studied by \cite{eva09} and the theoretical synchrotron spectra of their afterglows \citep{sar98}.
This discrepancy strongly indicates an absorbed afterglow, but no local dust model fits the data. There is no evidence of the 2175\,\AA{} graphite bump, which would be located in the $\Gg$ band based on the fixed redshift $z$ of the afterglow, excluding the \ac{MW} and \ac{LMC} dust models. The \ac{SMC} dust model in contrast features a strong ultraviolet extinction, which is not implied by the afterglow \ac{SED} either. Thus, a specific dust model without strong features or curvature similar to the model of an \ac{SB} is required in this case.
These considerations are strengthened by the contour plot shown in Fig.~\ref{fig_fit_aglow}, which signifies that an acceptable spectral index of $\bO^\mathrm{SB} \leq 1.2$ can be obtained by a moderate to high intrinsic extinction of $\AV^\mathrm{SB} \geq 0.6\,\mathrm{mag}$ within the 68\% confidence level.

\begin{figure*}[th]
\sidecaption
\includegraphics[width=120mm]{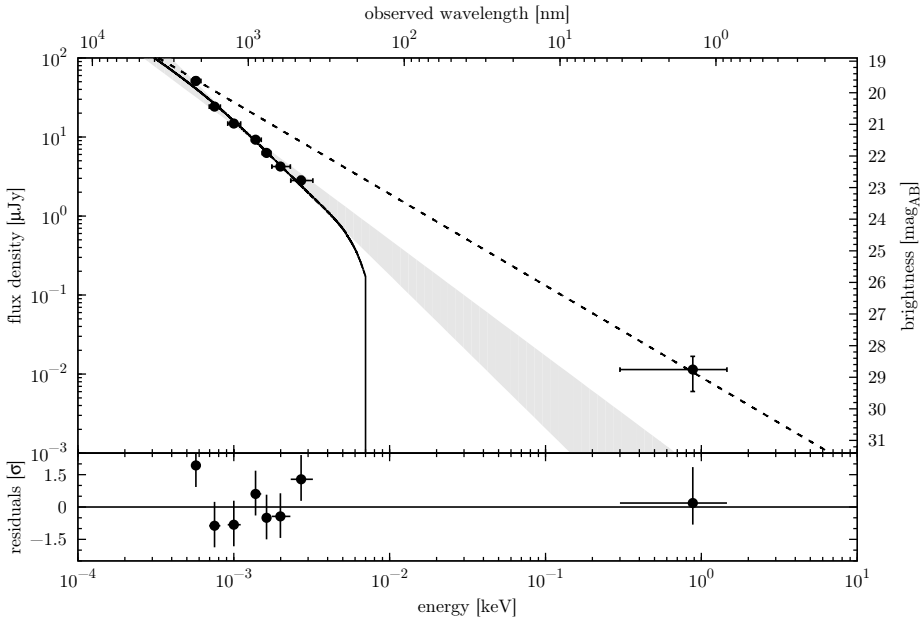} 
\caption[fc]{Broad-band spectral energy distribution of the afterglow of GRB\,071028B including GROND and \Swift/XRT data points. Based on the dust model of a starburst (SB), the shaded area represents the extrapolation of the optical and near-infrared data points with the best-fit spectral index of $\bO^\mathrm{SB} = 1.85\,^{+0.09}_{-0.38}$ and without dust. Using \textit{Minuit}, the data points are best fit by a single power law with a spectral index of $\bOX^\mathrm{SB} = 1.16 \pm 0.08$ (dashed line) in combination with the dust model of an SB with an intrinsic extinction of $\AV^\mathrm{SB} = (0.70 \pm 0.11)\,\mathrm{mag}$ (solid line).}
\label{fig_sedbb2}
\end{figure*}

The broad-band \ac{SED} of the afterglow including \ac{GROND} and \Swift/XRT data points is shown in Fig.~\ref{fig_sedbb2}.
Here, the X-ray data point taken at \TOp{2.7\,d} is extrapolated backwards to the time of the \ac{GROND} observations, assuming a normal-decay phase with a mean temporal index of $\aX = 1.2 \pm 0.2$ \citep{eva09}.
The systematic uncertainty of the backward extrapolation, based on the maximum and minimum values of $\aX$, is added quadratically to the statistical measurement error of the X-ray data point.
This approach is preferred to the forward extrapolation of the optical and near-infrared \ac{SED}, since no strong spectral evolution is expected from late X-ray observations.

Shown in Fig.~\ref{fig_sedbb2} is the extrapolation of the optical and near-infrared \ac{SED} with the best-fit spectral index of $\bO^\mathrm{SB} = 1.85\,^{+0.09}_{-0.38}$ and without dust, which significantly underpredicts the X-ray data point.
It is unlikely that strong flaring activity causes this excess in the X-ray flux,
since only seven (2\%) out of 287 \acp{GRB} detected by \Swift\ until the end of December 2007 exhibit late flares from $10^4$\,s after the trigger \citep{cur08}.
In this sample, the X-ray flares of only two ($<$ 1\%) \acp{GRB} show a strong relative flux variability of $\Delta F / F \gtrsim 10$, and the maximum value of $\Delta F / F \approx 70 \pm 40$ is estimated for GRB\,050916.
Assuming an equally strong flare for GRB\,071028B, the extrapolation of the optical and near-infrared \ac{SED} would still underpredict the X-ray data point by a factor of about three.
These findings also indicate that the afterglow is absorbed by significant amounts of dust.

We therefore utilised \textit{Minuit} for fitting a single power law in combination with the dust model of an \ac{SB} to the broad-band \ac{SED} of the afterglow.
Shown in Fig.~\ref{fig_sedbb2} is the best-fit solution with a goodness-of-fit of $\Xsqrdof = 6.86/5$, which yields a spectral index of $\bOX^\mathrm{SB} = 1.16 \pm 0.08$ and an intrinsic extinction of $\AV^\mathrm{SB} = (0.70 \pm 0.11)\,\mathrm{mag}$ for the afterglow.

\ac{GROND} systematically observed 128 \acp{GRB} from GRB\,070802 to the end of March 2010 \citep{gre10}. Thirty-nine long-duration \acp{GRB} out of this sample were observed less than 4\,hr after the trigger, and they exhibit an afterglow detected by \Swift/XRT. In the majority of these cases and for the first time, the intrinsic extinction $\AV$ of each afterglow was properly derived from a direct measurement of the respective optical and near-infrared \ac{SED}.
These findings indicate that the afterglow of GRB\,071028B is not exceptional, but among the most absorbed 12\% out of 33 long-duration \acp{GRB} with well known intrinsic extinctions.

While there has been evidence of a number of highly absorbed afterglows in recent years \citep[e.g.][]{jau08,kru08,tan08,pro09}, GRB\,071028B is special thanks to the relatively flat extinction curve. The only other \ac{GRB} so far with clear evidence of a flat extinction curve is GRB\,020813 \citep{sav04}.

\subsection{The energetics}

GRB\,071028B was observed by the \Swift/BAT in the energy range of 15\,keV -- 150\,keV, and the resulting prompt emission spectrum can be described well by a single power law with a photon index of $\varGamma = 1.45 \pm 0.25$ \citep{gcnr105}.
Applying this to the equation of \cite{sak09}, we calculate a value of $\Ep = (110\,^{+370}_{-60})\,\mathrm{keV}$ for the peak energy of the Band function \citep{ban93}. We note that the error estimate includes both the $1\sigma$ confidence level of the relation and the 90\% uncertainty of the photon index.
The normalisation parameter $A$ of the Band function is based on an energy fluence of $\varphi = \oof{(2.5 \pm 0.8)}{-7}\,\mathrm{erg}\,\mathrm{cm}^{-2}$, which is deduced from \Swift/BAT observations \citep{gcnr105}.
Completing the set of spectral parameters, we use $\alpha = -\varGamma$ and assume a standard value of $\beta = -2.5$ for the low and high energy index of the Band function, respectively \citep{pre00}.

We now adopt the energy range of 1\,keV -- 10\,MeV and apply the previously defined spectral parameters of $\alpha$, $\beta$, and $\Ep$ to the normalised Band function of the observer's frame. By integrating this over the blue-shifted energy range, an isotropic energy release of $\Erest = \oof{(1.4\,^{+2.4}_{-0.7})}{51}\,\mathrm{erg}$ in the rest frame is calculated. We note that the error estimate is based on the maximum and minimum value of each spectral parameter and the redshift.

In comparison, \cite{koc08} have studied the isotropic energy release of 63 long-duration \acp{GRB} detected by \Swift, and derive a median value of roughly $\Erest = \oof{(4.11\,^{+2.53}_{-0.54})}{52}\,\mathrm{erg}$. Hence, GRB\,071028B is approximately 30 times less energetic than this median of the \acp{GRB} with known redshifts, and is among the lowest 8\% of this sample.

\section{Summary and conclusions}

GRB\,071028B was observed by the \Swift/BAT, but no automated localisation was possible since this \ac{GRB} was located near the edge of the field-of-view.
Despite a late ground-based detection, we discovered the clearly fading afterglow with \ac{GROND}. Our follow-up observations revealed a constant and extended source next to the afterglow position, which we identified as the underlying host galaxy.

Based on accurate seven-channel photometry covering the optical to near-infrared wavelength range, we placed constraints on the characteristics of the afterglow and host galaxy.
Although previous studies have shown relatively low amounts of dust in \ac{GRB} environments, the afterglow of GRB\,071028B is moderately absorbed and requires a relatively flat extinction curve. In contrast, there is no evidence that the host galaxy is absorbed. These results indicate that the dust is local to the \ac{GRB} environment or highly unevenly distributed within the host galaxy, a conclusion already drawn for a few other \ac{GRB} lines-of-sight \citep{per09}.
We discussed the energetics further and concluded that this \ac{GRB} is among the lowest 8\% \citep{koc08}.
Observationally, the afterglow of GRB\,071028B at \TOp{2.0\,d} is of comparable brightness to the afterglows in the incomplete sample of 50 long-duration \acp{GRB} detected by \Swift\ until May 2009 and studied by \cite{kan10}. Accounting for the observational bias and assuming that this sample is reasonably complete at the bright end of the distribution, the afterglow of GRB\,071028B would be among the optically brightest 9\% out of 370 \acp{GRB} with X-ray afterglows detected by \Swift\ until May 2009. It also has a similar brightness to the afterglows of all \acp{GRB} detected by \Fermi/LAT \citep{bre10}.

\begin{acknowledgements}
We thank the referee for very helpful comments that allowed us to significantly improve the quality of the paper.
T.~Kr\"{u}hler acknowledges support by the \ac{DFG} cluster of excellence \lq Origin and Structure of the Universe\rq.
S.~Savaglio acknowledges support through project M.FE.A.Ext00003 of the \ac{MPG}.
The Ph.D. studies of F.~Olivares~E. are funded by the \ac{DAAD}.
P.~Schady acknowledges support through project SA 2001/2-1 of the \ac{DFG}.
S.~Klose and A.~Rossi acknowledge support by the \ac{DFG} through grant Kl 766/11-3.
A.~C.~Updike appreciates travel support through the \ac{MPE}.
Part of the funding for \ac{GROND} (for both hardware and personnel) was granted from the Leibniz-Prize to Prof. G.~Hasinger (\ac{DFG} grant HA 1850/28-1).
This work made use of data supplied by the United Kingdom \Swift\ Science Data Centre at the University of Leicester.
\end{acknowledgements}

\bibliographystyle{aa}
\bibliography{15318bib}

\end{document}